\documentclass[12pt,a4paper]{article}
\usepackage{amsmath,amssymb,url}
\usepackage{graphicx,tabularx}
\usepackage{array,dsfont}
\usepackage{thophys}

\makeatletter
\def\preprintno#1{\def\@preprintno{#1}}
\def\address#1{\def\@address{#1}}
\def\email#1#2{\thanks{\tt #1@{}#2}}
\def\titlenote#1{\def\@titlenote{#1}}
\def\abstract#1{\def\@abstract{#1}}
\renewcommand\abstractname{ABSTRACT}
\newlength\preprintnoskip
\newlength\abstractwidth
\@titlepagetrue
\renewcommand\maketitle{\begin{titlepage}%
  \let\footnotesize\small
  \hfill\parbox{\preprintnoskip}{%
  \begin{flushright}\@preprintno\end{flushright}}\hspace*{1cm}
  \vskip 60\p@
  \begin{center}%
    {\Large\bf\boldmath \@title \par}\vskip 1cm%
    {\sc\@author \par}\vskip 3mm%
    {\@address \par}%
    \vskip 2cm
    {\small\@titlenote \par}%
  \end{center}\par
  \@thanks
  \vfill
  \begin{center}%
    \parbox{\abstractwidth}{\centerline{\abstractname}%
    \vskip 3mm%
    \@abstract}
  \end{center}
  \end{titlepage}%
  \setcounter{footnote}{0}%
  \let\thanks\relax\let\maketitle\relax
  \gdef\@thanks{}\gdef\@author{}\gdef\@address{}%
  \gdef\@title{}\gdef\@abstract{}\gdef\@preprintno{}
}%
\def\@citex[#1]#2{\if@filesw\immediate\write\@auxout{\string\citation{#2}}\fi
  \def\@citea{}\@cite{\@for\@citeb:=#2\do
    {\@citea\def\@citea{,\penalty\@m}\@ifundefined
       {b@\@citeb}{{\bf ?}\@warning
       {Citation `\@citeb' on page \thepage \space undefined}}%
\hbox{\csname b@\@citeb\endcsname}}}{#1}}
\def\citerange{\@ifnextchar [{\@tempswatrue\@citexr}{\@tempswafalse\@citexr[]}}
\def\@citexr[#1]#2{\if@filesw\immediate\write\@auxout{\string\citation{#2}}\fi
  \def\@citea{}\@cite{\@for\@citeb:=#2\do
    {\@citea\def\@citea{--\penalty\@m}\@ifundefined
       {b@\@citeb}{{\bf ?}\@warning
       {Citation `\@citeb' on page \thepage \space undefined}}%
\hbox{\csname b@\@citeb\endcsname}}}{#1}}
\long\def\@makecaption#1#2{%
  \sbox\@tempboxa{#1: \emph{#2}}%
  \ifdim \wd\@tempboxa >\hsize
    #1: \emph{#2}\par
  \else
    \hbox to\hsize{\hfil\box\@tempboxa\hfil}%
  \fi
  \vskip\belowcaptionskip}
\def\fmslash{\@ifnextchar[{\fmsl@sh}{\fmsl@sh[0mu]}}
\def\fmsl@sh[#1]#2{%
  \mathchoice
    {\@fmsl@sh\displaystyle{#1}{#2}}%
    {\@fmsl@sh\textstyle{#1}{#2}}%
    {\@fmsl@sh\scriptstyle{#1}{#2}}%
    {\@fmsl@sh\scriptscriptstyle{#1}{#2}}}
\def\@fmsl@sh#1#2#3{\m@th\ooalign{$\hfil#1\mkern#2/\hfil$\crcr$#1#3$}}
\makeatother

\newcommand\ltap{\
  \raise.3ex\hbox{$<$\kern-.75em\lower1ex\hbox{$\sim$}}\ }
\newcommand\gtap{\
  \raise.3ex\hbox{$>$\kern-.75em\lower1ex\hbox{$\sim$}}\ }

\newcommand\simge{\mathrel{%
   \rlap{\raise 0.511ex \hbox{$>$}}{\lower 0.511ex \hbox{$\sim$}}}}
\newcommand\simle{\mathrel{
   \rlap{\raise 0.511ex \hbox{$<$}}{\lower 0.511ex \hbox{$\sim$}}}}

\newcommand\be{\begin{equation}}
\newcommand\ee{\end{equation}}
\newcommand\bea{\begin{eqnarray}}
\newcommand\eea{\end{eqnarray}}
\newcommand\ba{\begin{array}}
\newcommand\ea{\end{array}}

\def\bq{\begin{equation}}
\def\eq{\end{equation}}
\def\ba{\begin{eqnarray}}
\def\ea{\end{eqnarray}}

\newcommand{\LL}{\mathcal{L}}

\newcommand{\GeV}{{\ensuremath\rm GeV}}

\newcommand{\ii}{\mathrm{i}}

\newcommand{\vW}{\mathbf{W}}

\newcommand{\vH}{\mathbf{H}}
\newcommand{\tr}[1]{\operatorname{tr}\left[#1\right]}
\newcommand{\vB}{\mathbf{B}}
\newcommand{\vD}{\mathbf{D}}

\newcommand{\amp}{\mathcal{A}}
\newcommand{\WHIZARD}{\texttt{WHIZARD}}

\begin{document}

\date{\today}

\preprintno{DESY 14-044\\SI-HEP-2014-06}

\title{Simplified Models for Vector Boson Scattering \\at ILC and CLIC}

\author{J.~Reuter\email{juergen.reuter}{desy.de}$^a$, 
  W.~Kilian\email{kilian}{physik.uni-siegen.de}$^b$,
  M.~Sekulla\email{sekulla}{physik.uni-siegen.de}$^b$}

\address{\it%
$^a$DESY Theory Group, \\
  Notkestr. 85, D-22607 Hamburg, Germany
\\[.5\baselineskip]
$^b$University of Siegen, Physics Department, \\
  Walter-Flex-Str. 3, D-57068 Siegen, Germany
}

\titlenote{%
  Talk presented 
  at the International Workshop on Future Linear Colliders (LCWS13),
  \\
  Tokyo, Japan, 11-15 November 2013
}

\abstract{
Quasi-elastic scattering of the vector bosons $W$ and $Z$ is a
sensitive probe of the details of electroweak symmetry breaking, and a
key process at future lepton colliders.  We discuss the
limitations of a model-independent effective-theory approach and describe
the extension to a class of Simplified Models that is applicable to
all energies in a quantitative way, and enables realistic Monte-Carlo
simulations.  The framework has been implemented in the Monte-Carlo event
generator \WHIZARD.
}

\maketitle


\section{Introduction}

After the discovery of a Higgs-like particle at the LHC, and without
clear signals of new physics, the focus of particle physics has moved
towards a detailed study of the electroweak symmetry breaking
sector~\cite{Baak:2013fwa}.  Regarding observable processes at
colliders, the most striking effect of the Higgs boson is a delicate
cancellation in the scattering of on-shell electroweak vector bosons
(VBS).  This cancellation would be spoiled by any deviation from the
minimal Standard Model (SM).  Therefore, the study of high-energy
electroweak vector-boson scattering is an important part of the
physics program of a future lepton
collider~\cite{Accomando:1997wt,Djouadi:2007ik,Linssen:2012hp,Baer:2013cma}.

At high energies, the electroweak gauge bosons effectively decompose
into four gauge bosons $W^{\pm,0}_\mu,B^0_\mu$ with two transverse
polarization components, and three scalar Goldstone bosons $w^{\pm,0}$
which correspond to the longitudinal polarization components of $W$
and $Z$.  Consequently, the high-energy scattering amplitudes of $W$
and $Z$ bosons can be broken down into scattering of Goldstone bosons,
gauge bosons, and a mixed mode, respectively.

Gauge-boson interactions respect unitarity.  The calculated scattering
amplitudes of scalar Goldstone bosons are unitary only if they are
part of a gauge multiplet which transforms linearly.  Without any
Higgs boson, there is no linear realization, and therefore the
calculated scattering amplitudes of Goldstone bosons rise with energy,
in violation of the unitarity limit~\cite{Lee:1977yc,Lee:1977eg}.
Conversely, the SM Higgs boson implements a linear realization, so it
cancels this rise.  This is also true for any other Higgs-sector
incarnation in a linear representation.  Actually, with the measured
small Higgs mass of $m_H=125\;\GeV$, the asymptotic value of the
Goldstone-boson scattering amplitude is rather suppressed, and the
quasi-elastic scattering of $W$ and $Z$ bosons in the SM is dominated
by the transversal degrees of freedom.

The ILC and CLIC colliders will probe the VBS processes
\begin{equation}\label{VV}
  V V \to V V \qquad\text{with $V=W^+,W^-,Z$}
\end{equation}
embedded in the class of processes
\begin{equation}\label{ee}
  e^- e^+ \to \ell \ell' V V \qquad\text{with
    $\ell,\ell'=e^{\pm},\nu$},
\end{equation}
where the initial vector bosons of process~(\ref{VV}) are represented,
in (\ref{ee}), by virtual vector bosons radiated from the incoming
electron and positron,
respectively~\cite{Barger:1995cn,Boos:1997gw,Accomando:1997wt}.  The
virtual particles are space-like with a typical off-shellness
$p_T(\ell)\sim m_W$.  The final-state vector bosons decay into leptons
or quarks and are off-shell only by $\Gamma_{W,Z}$.

The ILC/CLIC environment essentially fixes the total energy of the
leptonic process.  This results in a spectrum of the $VV$ invariant
mass $m_{VV}$, which falls down with increasing value of
$m_{VV}$~\cite{Chanowitz:1984ne,Kane:1984bb,Dawson:1984gx}.  This
situation is more favorable than at the LHC, where the analogous
processes are further suppressed by the steeply falling parton
distribution functions.

The physical question is whether the SM is correct, as an effective
theory, up to the energy scale which is accessible in the collider
experiment.  VBS processes are especially important, because they
directly probe the Higgs sector, and because the SM prediction is
strongly suppressed.  Any small deviation can result in a large
effect, if sufficiently high energies can be reached.

\section{Effective Theory}

Traditionally, deviations from the SM prediction are parameterized by
an effective Lagrangian which contains, in addition to the
renormalizable (SM) part, a series of higher-dimensional effective
operators with prefactors of order $1/\Lambda^{d-4}$.  Here, $d$ is the
dimension of the operator.  The scale $\Lambda$ is unknown.  If the
operator is the result of tree-level exchange of a new particle,
$\Lambda$ is of the order of the particle mass $m$.  If it is the result
of a radiative correction, $\Lambda$ is rather of the order $4\pi m$.
If it is the result of a strong interaction, e.g., a composite nature
of the SM particles, $\Lambda$ indicates the compositeness scale.

In the effective Lagrangian of gauge bosons, Goldstone bosons, and
Higgs, the leading dimension $d$ is six.  If we only consider
operators which do not affect vector-boson self-energies or trilinear
couplings, they start at $d=8$.  Such interactions are generated by
tree-level exchange of new particles, if they exist.

We write the SM Lagrangian, excluding fermions, in the form
\begin{align}
  \LL_{\text{min}}=&-\frac{1}{2}\tr{\vW_{\mu\nu}\vW^{\mu\nu}}
		-\frac{1}{2}\tr{\vB_{\mu\nu}\vB^{\mu\nu}} \\
		&-\frac{1}{2}\tr{ \left ( \vD_\mu \vH \right )^\dagger
		 \vD^\mu \vH }
		- \frac{\mu^2}{4}\tr{\vH^\dagger \vH}
		+\frac{\lambda}{16}\left( \tr{\vH^\dagger \vH} \right)^2,
\end{align}
where we use the notation
\begin{align}
  \vD_\mu \vH 
  &= \partial_\mu \vH + \ii g \vW_\mu \vH - \ii g^\prime \vH \vB_\mu 
  \\
  \vW_{\mu\nu}
  &=\partial_\mu \vW_\nu - \partial_\nu \vW_\mu 
  + \ii g \left [ \vW_\mu , \vW_\nu \right ] 
  \\
  \vB_{\mu\nu}
  &=\partial_\mu \vB_\nu - \partial_\nu \vB_\mu
\end{align}
with the fields
\begin{align}
  \vW_{\mu} &= W_\mu^a \frac{\tau^a}{2}, &
  \vB_{\mu} &= B_\mu^a \frac{\tau^3}{2}, &
  \vH &=
  \frac 1 2 
  \begin{pmatrix}
    v + h -\ii w^3 & -\ii \sqrt{2} w^+ \\
    -\ii \sqrt{2} w^- & v + h + \ii w^3  \\
  \end{pmatrix}.
\end{align}
The Higgs field, which includes the Goldstone bosons, is represented here
as a $2\times 2$ hermitian matrix\footnote{The matrix form allows for simply
  relating the higher-dimensional operators to corresponding operators
  in the no-Higgs
  scenario~\cite{Appelquist:1980vg,Longhitano:1980iz,Longhitano:1980tm,Appelquist:1993ka}.
  Alternatively, by breaking down the matrix into columns, we can
  relate to the operators in Higgs-doublet
  notation~\cite{Buchmuller:1985jz,Hagiwara:1993ck,Eboli:2006wa,Degrande:2013rea}}.

The possible anomalous effective interactions which directly influence
VBS processes contain higher powers and higher derivatives of these
building blocks, for instance the two dimension-eight operators
\begin{alignat*}{3}
  \LL_{S,0} &= & & 
  F_{S,0} &&\tr{(\vD_\mu \vH)^\dagger(\vD_\nu \vH)} \times
  \tr{(\vD^\mu \vH)^\dagger(\vD^\nu \vH)} \\
  \LL_{S,1} &= & & 
  F_{S,1} &&\tr{(\vD^\mu \vH)^\dagger(\vD_\mu \vH)} \times
  \tr{(\vD^\nu \vH)^\dagger(\vD_\nu \vH)} 
\end{alignat*}
which modify the pure Goldstone-boson part of VBS.

Augmenting the SM Lagrangian by a complete (and minimal) set of
additional operators with unknown coefficients, we can model the
behavior of the collider processes in a certain energy range,
represented by $m_{VV}$.  Experiment would determine or limit the
values of those
coefficients~\cite{Accomando:1997wt,Beyer:2006hx,Djouadi:2007ik,Baak:2013fwa}.

In VBS processes, we are in the special situation that the interesting
operators have dimension eight, and the SM contribution is small.
This is in contrast to the analogous situation in $W$ pair
production~\cite{Hagiwara:1993ck,Degrande:2012wf} or VBS without light
Higgs~\cite{Boos:1997gw}.  Just slightly above the point where the
anomalous effect becomes detectable, the squared amplitude rises
steeply ($\sim s^4$) in relation to the SM reference distribution, and
crosses the unitarity bound.  This severely limits the use of the
effective theory.

\section{Beyond the Limit}

This situation is rather unsatisfactory.  In essence, there is no
model-independent description of high-energy VBS that is both
different from the SM and phenomenologically useful.  It would be
possible to switch to a top-down description where all observables are
calculated within a specific model.  Unfortunately, while we know some
non-SM models of the Higgs sector such as multi-Higgs doublet models,
the really interesting cases of strong interactions or compositeness
where large effects are possible, are rather uncertain.

We therefore propose to model the physics of VBS by an improved
effective theory, as an attempt to make the effective theory
consistent with general principles of quantum theory, and to cover a
wide class of possible extensions into the energy range where no
definite predictions exist~\cite{Alboteanu:2008my,Reuter:2013gla}.

The first step is the unitarization of the low-energy effective theory.  Given
an arbitrary approximation to the true $S$ matrix, we may resum the scattering
in such a way that the result is unitary, without losing information or
introducing arbitrary new parameters.  A simple application is Dyson
resummation, which turns an intermediate on-shell particle into a Breit-Wigner
resonance.  This recipe, which is known as the K-matrix  \cite{Dicus:1989zm} or inverse-amplitude
unitarization procedure, is particularly simple if we are dealing with a $2\to
2$ scattering process.  If we diagonalize the $VV\to VV$ scattering amplitudes
in the high-energy limit, VBS satisfies this condition.  Given a correctly
normalized real-valued diagonal amplitude $\amp(s)$, the unitarized amplitude
reads
\begin{equation}
  \amp^K(s) = \amp(s) / (1 - \ii\amp(s)).
\end{equation}
If, as in VBS with anomalous contributions, $\amp(s)$ rises without bound for
increasing energy, the asymptotic value is always $\amp^K(\infty)= \ii$, which
is the maximum absolute value consistent with unitarity.  The unitarized
amplitude asymptotically saturates the unitarity limit.  For low energies
where $|\amp(s)|\ll 1$, it coincides with the effective-theory result.  The
coefficients of the higher-dimension operators thus describe the \emph{slope}
of the approach to saturation.

Of course, the real physics may behave in a completely different way -- as
long as the saturation limit is not exceeded.  In accordance with unitarity,
the elastic amplitude which must lie on the Argand circle $|\amp(s)-1/2|=1/2$,
may actually pass the point $\amp=\ii$.  This is a resonance, which can be
described by a mass and width parameter.

We therefore include resonances in the description.  In the Higgs-Goldstone
sector, the interacting particles $h,w^\pm,z$ are scalars and have definite
quantum numbers regarding weak isospin (custodial symmetry), so the scattering
is diagonalized in terms of spin-isospin eigenamplitudes $\amp_{IJ}(s)$, where
$I, J$ is in the range $0,1,2$.  Furthermore, with exact isospin, accessible
resonances have even values of $I+J$.  There are five possibilities: scalar
($I=0$ or $I=2$), vector ($I=J=1$), or tensor ($I=0$ or $I=2$).  This amounts
to five mass and five coupling parameters, which then determine the resonance
widths.

This discussion is not just academic -- at sub-$\GeV$ energies, QCD behaves in
exactly this way.  There, the $\rho$ vector resonance dominates scattering
amplitudes, but this need not be the case for electroweak interactions.  In
any case, it is reasonable to assume that this approach covers the dominant
effects both in weakly interacting scenarios (such as multi-Higgs models, a
special case where all resonances are scalar) and in the presence of strong
interactions and compositeness.

In summary, we propose to model the physics of vector boson by a {Simplified}
model of {Strong} interactions and {Compositeness}, which we will denote as
\textbf{SSC}.  At low energies, it smoothly approaches the SM, and at
intermediate energies below the first resonance (if any), it is equivalent to
the generic effective theory.  The ingredients are:
\begin{enumerate}
\item
  All particles of the SM, including the Higgs boson
\item
  The leading higher-dimensional operators, possibly restricted to those that
  affect the process in question, with arbitrary coefficients
\item
  The full set of resonances accessible in Goldstone- (and Higgs) scattering,
  parameterized by independent mass and coupling parameters
\end{enumerate}
We emphasize that the resonances, if integrated out, do contribute a
shift to the low-energy anomalous couplings.  We do not assume
that they are the only contribution.

\section{Implementation}

While the model is formulated in terms of the high-energy degrees of
freedom, it is straightforward to extend it to the full range of
energies~\cite{Alboteanu:2008my}.  The SM, effective operators, and
resonances, are expressed as a Lagrangian and can thus be evaluated in
terms of Feynman rules with off-shell particles.  The model is useful
as long as the off-shellness is small compared to the relevant energy
scale.  However, at low energies where this is not the case, the
unitarization corrections and the resonance effects disappear.  The
corrections due to unitarization are not expressible as a Lagrangian,
but they behave as well defined, Lorentz-invariant and gauge-covariant
form factors, so they can be extrapolated off-shell and included in
the calculation in a straightforward way.

The unitarization corrections violate crossing symmetry.  Thus, results
obtained for $VV\to VV$ processes cannot directly be translated
into predictions for the crossed process $V^*\to VVV$, or vice versa, as
it would be possible in the pure SM.

We have implemented this framework in the Monte-Carlo event generator
\WHIZARD~\cite{Kilian:2007gr,WHIZARD-LCWS13}.  In the upcoming version
2.2 of this generator, the relevant physics models are
\begin{description}
\item[SM]: The pure Standard Model as reference
\item[SM\_ac]:  Standard Model with anomalous interactions, as a consistent
  effective theory but violating unitarity
\item[SSC]: The Simplified Model with anomalous interactions, resonances, and
  K-matrix unitarization, as described above
\end{description}
For convenience, there is also the alternative model
\begin{description}
\item[AltH]: The Simplified Model without the light Higgs boson
\end{description}
This model allows for connecting to older VBS studies in the literature, where
the focus was on the no-Higgs scenario.

\section{Conclusions}

The SSC approach enables a reasonable evaluation of the sensitivity of
a future lepton collider, regarding the physics of high-energy VBS.
It should also be useful for the analysis of real data, and in
particular for combining ILC and CLIC with hadron collider results.
If non-SM effects are actually found, the model should be extended to
describe physics phenomenologically in more detail.  For instance, one
should allow for weak-isospin violating corrections and independent
couplings to transversal vector bosons.  The setup for the Higgs
sector will be embedded in the context of potential new physics in the
fermion and gauge sectors, and it should be connected with specific
models of the new interactions at a more fundamental level.

In the opposite, but not unlikely case that no deviations can be detected, the
traditional effective-theory approach loses its value for data analysis at the
energies that we want to access at the ILC and, in particular, at CLIC.
Instead, the SSC framework extends the effective theory and allows us to
quantify the precision with which the SM will be verified.
     

\section*{Acknowledgments}

JRR has been partially supported by the Strategic Alliance ``Physics
at the Terascale'' of the Helmholtz-Gemeinschaft. JRR and WK want to
thank the organizers for scheduling the conference in the amazing
Japanese autumn foliage season.





\end{document}